\documentclass[aps,prl,twocolumn,showpacs,showkeys,amsmath,amssymb,superscriptaddress]{revtex4}

\usepackage{amssymb,amsmath}
\usepackage{graphicx}
\usepackage{epsfig}

\def\Tr{\mbox{Tr}\,}

\newcommand{\ee}{\end{equation}}
\newcommand{\be}{\begin{equation}}
\newcommand{\bea}{\begin{eqnarray}}
\newcommand{\eea}{\end{eqnarray}}
\newcommand{\eu}{{\rm e}}
\newcommand{\ii}{{\rm i}}
\newcommand{\sign}{{\rm sgn}}
\newcommand{\de}{{\displaystyle\rm\mathstrut d}}

\def\XXint#1#2#3{{\setbox0=\hbox{$#1{#2#3}{\int}$}
     \vcenter{\hbox{$#2#3$}}\kern-.5\wd0}}

\begin{document}

\title{Horizon in Random Matrix Theory, Hawking Radiation and Flow of Cold Atoms}

\author{Fabio Franchini \& Vladimir E. Kravtsov, \\
The Abdus Salam ICTP, Strada Costiera 11, Trieste (Italy)}

\date{\today}

\begin{abstract}
We propose a Gaussian scalar field theory in a curved 2D metric with an event horizon as the low-energy effective theory for a weakly confined, invariant Random Matrix ensemble (RME). The presence of an event horizon naturally generates a bath of Hawking radiation, which introduces a finite temperature in the model in a non-trivial way. A similar mapping with a gravitational analogue model has been constructed for a Bose-Einstein condensate (BEC) pushed to flow at a velocity higher than its speed of sound, with Hawking radiation as sound waves propagating over the cold atoms. Our work suggests a three-fold connection between a moving BEC system, black-hole physics and unconventional RMEs with possible experimental applications.
\end{abstract}

\pacs{71.10.Pm, 72.15.Rn, 73.23.-b, 04.62.+v,  04.70.Dy, 03.75.Kk, 67.85.Hj}

\keywords{Random Matrix Theory, Luttinger Liquid, Hawking Radiation, Event Horizon, Cold Atoms, Bose-Einstein Condensate, QFT in Curved Space-time, Bosonization}

\maketitle

One of the most spectacular results in general relativity is the
prediction by Hawking \cite{Hawking74, Hawking75} of a thermal radiation whose
existence is only due to the presence of an event horizon near the
black hole singularity. Unfortunately, direct
experimental evidence of Hawking radiation is almost impossible
(see, for instance, \cite{balbinot06}). After the beautiful proposal
by Unruh \cite{unruh81}, there has been a growing interest in
searching for similar phenomena in the so called analogous
gravitational models \cite{visser05}. One of such models can be
realized in a stream of cold atoms with an abrupt change of the
stream velocity $v(x)$ at $x=0$. In such system, the event horizon
arises due to the fact that the phonon velocity $c$, which is larger
than stream velocity for $x<0$, abruptly gets smaller than $v$ for
$x>0$. Thus the entire region $x>0$ is analogous to the black hole
{\it interior} from where no particle can escape. However, unlike
the interior of black holes, the region $x>0$ is accessible to
measurements, in which case an experimentalist plays the role of a
super-observer. In this realization, Hawking radiation is
characterized by an intensity (or phonon density $n(x)$) which is
strongly {\it anti-correlated} not only at small distances
$|x-x'|\sim \bar{n}^{-1}$ but also at the mirror-reflected condition
$|x+x'|\sim \bar{n}^{-1}$, being exponentially small elsewhere
\cite{balbinot08}. Amazingly, exactly the same behavior was found
for eigenvalue density correlations (two-level correlation function
TLCF) in a certain exactly solvable random matrix ensemble (RME)
\cite{muttalib93, kravtsov95}.

In this letter we propose a connection between three very different areas of physics, namely Random Matrices, Quantum Gravity and Bose-Einstein Condensates (BEC). The relation between the latter two has been already established in \cite{balbinot08, carusotto08}. Here we focus on a non-standard invariant ensemble of Random Matrices described by a probability distribution $ P ( {\bf H} )=P({\bf UHU^{-1}})$ of the form
\be
   P ( {\bf H} )
   \propto \eu^{- \Tr V ( {\bf H} )} \; ,
   \qquad
   V ( E ) \stackrel{| E | \to \infty}{\simeq} {1 \over \kappa} \; \ln^2 | E | \; .
   \label{PofH}
\ee

In \cite{kravtsov95}, the asymptotic behavior of the TLCF for this weakly confined ensemble was evaluated and it was found that, beside the usual translational invariant part already found in \cite{muttalib93}, there was an {\it anomalous}, non-translational invariant component of the cluster function: \be
   Y_2^{\rm a} (x,x') ={\kappa^2 \over 4 \pi^2} \;
   {\sin^2 [\pi (x-x') ] \over \cosh^2 [ \kappa (x+x')/2] } , \;
   {\rm for} \; x \; x' < 0 ,
   \label{Y2a}
\ee
which corresponds to {\it repulsion} between the level at a point $x$ in the (unfolded) energy space and its {\it mirror image} at $x'$.

The main result of this letter is that the TLCF with this anomalous component
can be exactly reproduced within a scalar field theory (Luttinger liquid of levels \cite{kravtsov00}) in a curved space-time with an event horizon. The same kind of theory has already been shown in \cite{balbinot08} to be a good low-energy approximation for the BEC system pushed to move faster than its speed of sound, as it was confirmed by subsequent numerical simulations \cite{carusotto08}. The equivalence of the TLCFs for these theories strongly suggests some relation between them and unlock the possibility of extending techniques from one field to another. For instance, using the Luttinger Liquid description, we predict an oscillatory component in the TLCF of \cite{balbinot08}, which should be experimentally accessible for a BEC in the Tonks-Girardeau regime.

Due to the interdisciplinary nature of these subjects, we first briefly remind the reader of the different ingredients needed to establish these connections, so not to hinder the physical understanding behind the formulae.

{\it Hawking Radiation in a BEC: }
When a BEC fluid is pushed to move faster than its speed of sound, a {\it sonic black hole} with its event horizon is generated. This horizon separates a region where elementary excitations (phonons) can propagate along the fluid flow or against it and a region where the upstream propagation is suppressed, because the fluid velocity is so high that it drags any sound wave with it.
Physically, it is simpler to modulate the local sound velocity $c$, while keeping constant the density $\rho_0$ and velocity $v_0$ of the condensate \cite{balbinot08}. One can then design the set-up to have $c_{\rm l} < v_0 < c_{\rm r}$, where $c_{\rm l,r}$ are the sound velocities to the left/right of the horizon, where $c = v_0$.

The fact that the boundary between the sub- and super-sonic regions of a fluid constitutes a horizon is relatively intuitive. Much less obvious \cite{balbinot06} is the fact that the mere existence of a horizon kinematically implies the generation of Hawking radiation emanating from it \cite{unruh81}. {\it Analogue gravitational models} like this one are a topic of growing interest \cite{visser05}, since ``table-top'' experiments can be used to test predictions of quantum gravity and help in resolving some of its issues \cite{volovik}. However, the experimental detection of Hawking radiation has been so far impeded by its small temperature and intensity. The novel idea put forward in \cite{balbinot08} has been to investigate the TLCF as a signature of Hawking radiation, as pairs of phonons generated close to the horizon will remain highly entangled as they travel in opposite directions in the sub- and super-sonic region.

Sound waves propagate according to a d'Alembertian (a Laplace-Beltrami equation in a curved metric). The fluid is 3-dimensional, but since the relevant degree of freedom is only the one along the fluid motion, two spatial dimensions can be traced out and the effective dynamics is $1+1$.
In a flat space, the fundamental Green's function is $G(u) \propto - {1 \over 2} \ln \left( u^+ u^- \right)$, where $u^\pm \equiv t \pm \int {{\small {\rm d}} x \over c \mp v_0}$ are the light-cone coordinates.
The phase field for the condensate propagates according to this correlator,
the number density field being related to it by a differential operator
(see \cite{balbinot08} for the details).

The key element in \cite{balbinot08} is that, when a sonic black-hole horizon is created, the outgoing modes are exponentially redshifted in frequency and they propagate according to the effective light-cone coordinate
\be
   \tilde{u}^- \equiv \pm {1 \over \kappa} \; \eu^{-\kappa \; u^-} ,
   \label{umdef}
\ee
where $\kappa$ is a parameter known as the surface gravity at the horizon.
The ingoing modes propagate normally according to $u^+$.
Taking (\ref{umdef}) into account for the propagator, the two-point density-density correlation function can be calculated, yielding a translational invariant as well as an anomalous contribution for $x \; x' < 0$ \cite{balbinot08}:
\be
   Y_{2, {\rm BEC}}^{\rm a} (x,x') \propto
   \cosh^{-2} \left[ {\kappa \over 2}
   \left( {x \over c_{\rm r} - v_0} + {x' \over v_0 - c_{\rm l}} \right) \right] .
   \label{y2ap}
\ee
This correlator shows a characteristic off-diagonal peak connecting the sub- and super-sonic region of the fluid. In \cite{carusotto08}, this low-energy, field-theory based prediction has been tested against microscopic, ab-initio numerical simulation of the system showing an almost perfect agreement. Please notice that selecting $c_{\rm r, l} = v_0 \pm v_0/2$, (\ref{y2ap}) correctly reproduces the denominator in (\ref{Y2a}).

{\it The weakly confined RME: }
One exactly solvable RME with a confinement behaving like (\ref{PofH}) is
\be
   V(E) = \sum_{n=0}^\infty \ln \left[ 1 + 2 q^{n+1} \cosh (2 \chi) + q^{2n+2} \right] ,
   \label{VE}
\ee
with $E \equiv \sinh \chi$, $q \equiv \eu^{-\kappa}$ and $\kappa >0$
\cite{muttalib93}. The model has been solved using appropriate orthogonal
polynomials (q-deformed Hermite polynomials) and other methods and various
quantities have been calculated in the literature. One can introduce a change of variable that makes the mean level density $\rho (E) \equiv \Tr \left\{ \delta \left( E - {\bf H} \right) \right\}$ constant, i.e. $\langle \tilde{\rho} (x) \rangle \equiv \langle \rho (E_x) \rangle \; {\de E_x \over \de x} = 1$ . One of the most important peculiarities of this model is that the variable performing this unfolding in the bulk of the spectrum behaves exponentially \cite{kravtsov95}:
\be
   {\de x \over \de E } = {1 \over \kappa |E|} \qquad \Rightarrow
   \qquad E_x = \lambda \; \eu^{\kappa |x|} \; \sign (x) \, ,
   \label{Exdef}
\ee
where $\lambda$ is an (uninfluential) constant of integration.

The TLCF
is defined as
\be
   Y_2 (x,x') \equiv \delta ( x - x' ) -
   { \langle \rho (E_x) \rho (E_{x'}) \rangle -
   \langle \rho (E_x) \rangle \langle \rho (E_{x'}) \rangle \over
   \langle \rho (E_x) \rangle \langle \rho (E_{x'}) \rangle } .
   \nonumber
\ee
For $p \equiv \eu^{-2 \pi^2 / \kappa} \ll 1$ a semiclassical analysis can
be carried out \cite{kravtsov95} and one finds that the asymptotic behavior of the two-point correlation function is $Y_2 (x,x') = Y_2^n (x-x') \; \theta (x \; x') + Y_2^a (x,x') \; \theta (- \; x \; x') \;$ , where $\theta (x)$ is the traditional step function. The normal translational invariant part of the correlation function is \cite{muttalib93}
\be
   Y_2^n (x-x') = {\kappa^2 \over 4 \pi^2} \;
   {\sin^2 [\pi (x-x') ] \over \sinh^2 [ \kappa (x-x')/2] } , \;
   {\rm for} \; x \; x' > 0 ,
   \label{Y2n}
\ee while the anomalous (non-translational invariant) part is given by (\ref{Y2a}). As the anomalous component is the most interesting one, we do not concentrate on the normal component (\ref{Y2n}), even if it is also common to all theories considered in this paper. The main difference between (\ref{Y2a}) and (\ref{y2ap}) is the oscillatory part at the numerator of (\ref{Y2a}), the derivation of which is the focus of the next section.

{\it RME and Luttinger liquid: }
If we interpret the energy levels of a random matrix as the coordinates of a system of particles, the distribution of eigenvalues can be thought of as the
equilibrium configuration of such a system of {\it quantum interacting} particles in 1D, the phenomenon of level repulsion being encoded in their fermionic statistics and in the form of interaction. In this description, the probability distribution (\ref{PofH}) becomes a confining potential for the particles.  In condensed matter, it is well known that the low-energy behavior of a 1D system is not given by a Fermi Liquid theory, but instead by the so called {\it Luttinger Liquid} (LL).

The Luttinger Liquid paradigm stems from the fact that in 1D a system responds as a whole to any perturbation because particles cannot go around each others. Therefore, low-energy excitations have an intrinsic sound-wave nature and the system can be described in generality using only two parameters $K$ and $c$ within a quadratic action for a bosonic {\it displacement} field $\Phi(x,t)$:
\be
   {\cal S} [\Phi] = {\hbar \over 2 \pi K} \int_0^{1/T} \de \tau
   \int_{-\infty}^\infty \de x \;
   \left[ {1 \over c} \left( \partial_\tau \Phi \right)^2
   + c \left( \partial_x \Phi \right)^2 \right] \; .
   \label{T-Lutt}
\ee
Contrary to Fermi Liquid theory, the Luttinger Liquid description is valid for systems with fermionic statistics as well as for bosonic systems, although the details of the mapping are of course different \cite{giamarchi}.
In (\ref{T-Lutt}), $c$ is the wave (sound) velocity (we will set $\hbar = c=1$ henceforth) and the parameter $K$ encodes all the effective interactions of the original model: $K \to \infty$ for free bosons and decreases as the repulsive interaction increases, while for fermions $K>1$ corresponds to attraction, $K<1$ to repulsion and $K=1$ means free fermions. In this low energy approximation, the bosonic theory is quadratic and therefore the fundamental Green's functions is $G(x) \propto - {1 \over 2} \ln \left( x^2 \right)$.

The particle density operator can be written in terms of the displacement operator as \cite{giamarchi}
\be
   \rho (x,\tau) = \rho_0 - {1 \over \pi} \; \partial_x \Phi
   + {A_K \over \pi} \cos \left[ 2 \pi \rho_0  x - 2 \Phi \right]
   + \ldots \; ,
   \label{rhoexp}
\ee where $\rho_0$ is the background density -- $\rho_0=1$ in our RME after unfolding (\ref{Exdef}) -- and higher harmonics terms arise from interference effect ({\it Umklapp} processes mixing left and right movers in the fermionic language). $A_K$ is a constant that depends on the short-distance behavior.

We propose that the low-energy effective theory for the {\it invariant} weakly confined RME defined by (\ref{PofH}) is given by the {\it ground state} of a LL in a curved metric, i.e. by a bosonic field whose dynamic is dictated by the action
\be
   {\cal S} [\Phi] = {1 \over 2 \pi K} \int \de^2 \xi \sqrt{ g(\xi)}
   g^{\mu \nu} \partial_\mu \Phi \partial_\nu \Phi \; ,
   \label{CurvedS}
\ee where $g \equiv \det g^{\mu \nu}$ is the metric, i.e. $\de s^2 =
g_{\mu \nu} \de \xi^\mu \de \xi^\nu$. In this case the parameter $\kappa$ enters in the space-time metric $g$ rather than directly as a temperature $T$ like in (\ref{T-Lutt}). As we are interested in the low-energy asymptotics of the theory, we cannot specify the metric uniquely, but its causal structure will suffice.

In the construction, we are guided by the work done for the supersonic BEC \cite{balbinot08}. We need a horizon, since this causes an exponential redshift in the modes, which leads to the appearance of Hawking radiation and of an anomalous two-point correlation function like (\ref{Y2a}). However, the metric used in \cite{balbinot08} is physically motivated by the BEC setup, where time reversal invariance is explicitly broken, hence only the out-going modes are exponentially red-shifted. For the RME we will preserve T-reversal and therefore we will not have a radiation, but a bath with a finite temperature $T$.

In \cite{kravtsov00}, it has been shown that the traditional Luttinger liquid phenomenology (\ref{T-Lutt},\ref{rhoexp}) with $T\propto \kappa$ correctly reproduces the TLCF (\ref{Y2n}) of a {\it non-invariant} RME, which is a one-parameter deformation of the classic Wigner-Dyson RME, leading to the translational-invariant result (\ref{Y2n}). It is remarkable that (\ref{Y2n}) is common to both systems, while to obtain the anomalous part (\ref{Y2a}) we need to introduce the temperature in a non-trivial way through the metric (\ref{CurvedS}) and a horizon.

Let us then start with the {\it Rindler} line element
\be
  \de s^2 = - y^2 \; \de t^2 + {1 \over \kappa^2} \; \de y^2
  = - y^2 \; \de u^+ \; \de u^- ,
  \label{rindlermetric}
\ee
with $u^\pm \equiv t \pm {1 \over \kappa} \ln |y|$, which is arguably the simplest $1+1$ metric with a horizon and it can be used as a model to capture the universal features of many physical systems (for instance, as the dimensional reduction of the Schwarzschild black-hole solution -in Kruskal coordinates- or as the effective metric observed by a uniformly accelerated observer in flat background \cite{birreldavies}).

We can remove the singularity of the metric (\ref{rindlermetric}) at the horizon $y=0$ by introducing a new set of null (light-cone) coordinates:
\be
   \bar{u}^\pm \equiv \pm \; {1 \over \kappa} \; \eu^{\pm \kappa u^\pm}
   \sign (y) ,
   \label{rindler}
\ee
in terms of which we have a flat Minkowski space
\be
   \de s^2 =-\de \bar{u}^+ \; \de \bar{u}^- =-\de \bar{t}^2 + \de \bar{x}^2 ,
   \label{minkowski}
\ee
where $\bar{u}^\pm \equiv \bar{t} \pm \bar{x}$ (the plus/minus refers to left/right moving modes, i.e. analytic/antianalytic sectors).

Motivated by (\ref{VE},\ref{Exdef}), in (\ref{rindlermetric}) we chose
$y \equiv \sinh (\kappa x)$:
\bea
   && \de s^2 = - \sinh^2 (\kappa x) \; \de t^2
   + \cosh^2 (\kappa x) \; \de x^2 ,
   \label{metric} \\
   && \bar{x} = {1 \over \kappa} \sinh (\kappa x) \cosh (\kappa t) , \;
   \bar{t} = {1 \over \kappa} \sinh (\kappa x) \sinh (\kappa t) . \; \qquad
   \label{xtbar}
\eea
Lines of constant $t$ are straight lines through the origin, while curves of constant $x$ correspond to hyperbolae in the flat Minkowski space (\ref{minkowski}). It is known that hyperbolae are world lines of uniformly accelerated observers with acceleration $a=\kappa \; \sinh^{-1} (\kappa x)$ and indeed the set of coordinates (\ref{xtbar}) is a modified version of those of the ``traditional'' accelerated observer \cite{birreldavies} and coincide with them far from the origin:
\be
    \bar{x} \sim {\eu^{\kappa |x|} \over 2 \kappa} \cosh (\kappa t) \; \sign (x) , \;
    \bar{t} \sim {\eu^{\kappa |x|} \over 2 \kappa} \sinh (\kappa t) \; \sign (x) .
    \label{xt}
\ee
In fact, the vacuum state of the Minkowski space is seen as a thermal state in (\ref{metric}): an inertial observer would feel a black body ({\it Hawking}) radiation with temperature $T = \kappa / (2 \pi k_B)$. This can be seen by noticing the periodicity in the imaginary time of (\ref{xt}). But it can be shown more accurately by carefully decomposing the fields in modes in the two metrics and calculating their overlap \cite{birreldavies}.

We can now evaluate the density-density correlation function for the weakly confined RME using (\ref{rhoexp}):
\bea
  \label{Y2bos}
  Y_2 & = & -{1 \over \pi^2} \langle \partial_x \Phi (x) \partial_{x'} \Phi (x') \rangle \\
  && \quad - {A_K^2 \over 2 \pi^2} \cos (2 \pi (x-x')) \langle \eu^{\ii 2 \Phi (x)} \eu^{-\ii 2 \Phi(x')} \rangle + \ldots \; .
  \nonumber
\eea The non-oscillating part is simply
\be
   \langle \partial_x \Phi (x) \partial_{x'} \Phi (x') \rangle = - {K \over 2} \; \partial_x \partial_{x'} G(x,x') \; ,
   \label{nonosc}
\ee
and for a Gaussian theory we have
\be
   \langle \eu^{\ii \alpha \Phi (x)} \eu^{-\ii \alpha \Phi(x')} \rangle
   = \eu^{-{\alpha^2 \over 2 } \langle \left[ \Phi(x) - \Phi(x') \right]^2 \rangle } = \eu^{-{\alpha^2 \over 2} K G(x,x')} .
   \label{vertex}
\ee
The (equal-time) Green's function in the coordinates given by (\ref{rindler}) -- or (\ref{xt}) since we are interested in the asymptotics far from the origin-- is
\be
   G (x,x') \stackrel{|x|,|x'| \gg 1}{\simeq}
            \left\{ \begin{matrix}
                      \ln  \left[ {2 \over \kappa}
                      \sinh { \kappa (x-x') \over 2 } \right] , & \; x \; x'>0 \cr
                      \ln \left[ {2 \over \kappa}
                      \cosh { \kappa (x+x') \over 2 } \right] , & \; x \; x'<0 \cr
                      \end{matrix}
   \right. \; .
   \nonumber
\ee
Combining this with (\ref{Y2bos}--\ref{vertex}), it is straightforward to see that (\ref{Y2n}, \ref{Y2a}) are reproduced with $K=1$ (and $A_K=1$~\cite{giamarchi}).

{\it Conclusions:} We have shown that the simplest 2D metric with horizons yields the anomalous two-point function (\ref{Y2a}) characterizing the weakly confined RME and which was also found for a supersonic BEC (\ref{y2ap}). This suggests that all these theories share the same topological nature. This anomalous correlator is just another aspect of the emission of Hawking radiation at the horizons.

Eq. (\ref{rindler}) and (\ref{umdef}) are at the heart of the equivalence between these models. However, in (\ref{umdef}) only outgoing modes are exponentially redshifted, since the BEC system is not time-reversal invariant (the horizon is created at a finite moment in time). This means that the Hawking radiation is an actual flow of particles coming from the horizon. Instead, the metric we constructed for the RME is time-reversal invariant and therefore there is no radiation, but a thermal bath in equilibrium (Hartle-Hawking effect \cite{birreldavies}). This is required since both modes have to be redshifted in (\ref{rindler}) in order to give the correct prefactor in (\ref{Y2a}) compared to (\ref{Y2n}).

Two final remarks: (i) The Luttinger Liquid description for 1D systems (both fermionic and bosonic) directly leads to the emergence of oscillatory terms like those found in (\ref{Y2a},\ref{Y2n}). In our model, since $K=1$, these oscillations decay as fast as the leading term. These contributions were not present in the BEC result (\ref{y2ap}) \cite{balbinot08}, since one needs to employ a proper regularization scheme, like the one provided by the Luttinger Liquid paradigm, to calculate next-to-leading corrections. However for $K>1$ they would be suppressed according to (\ref{vertex}), but for a cold atomic system in the Tonks-Girardeau regime these oscillations should be observable. (ii) Both the invariant RME (\ref{PofH}) we studied here and the non-invariant, Power-Banded RME considered in \cite{kravtsov00} seem to be described by a thermal Luttinger Liquid theory and share the translational invariant TLCF (\ref{Y2n}). However, temperature in the invariant case is introduced in a non-trivial way through the Hawking effect, which leads to the anomalous component (\ref{Y2a}) as an indication of a non-trivial topology. It has been argued that the weakly confined invariant ensemble (\ref{PofH}) spontaneously breaks its rotational invariance \cite{kravtsov95,pato00} and would share many similarities with the system of \cite{kravtsov00}. If this was the case, one could use the plethora of analytical techniques available for invariant models to study a non-invariant system. Our work provides new evidences of the connection between these two RMEs, but leaves many questions still unanswered.

A microscopical derivation of the low-energy effective theory for the weakly confined RME could help in clarifying this latter point and explaining the role of topology. Moreover, what the Hawking radiation represents should be understood in the original RME model as well as the interpretation of the time coordinate (and the temperature) of the bosonic model, which in \cite{kravtsov00} was identified with the magnetic field threading the system. We should also remember that the RME (\ref{VE}) is an exactly solvable system and therefore its interpretation as a gravitational analogue model has a lot to offer, for instance in addressing the so called transplanckian problem.
All these points will be the focus of our next efforts.

{\it Acknowledgements:} We thank R. Balbinot, I. Carusotto and S. Fagnocchi for discussions on their work and comments on ours and A.A. Nersesyan for his help.

\vskip -0.5cm

\end{document}